\documentstyle[12pt]{article}
\begin{document}

\def\k{\kappa}
\def\d{\delta}
\def\vac{|0\rangle}
\def\S{\Sigma}
\def\H{{\cal H}}
\def\tr{{\rm tr}}
\def\O{\Omega}

\begin{center}
{\Large \bf Black Hole Entropy and Induced Gravity}\\
\vskip .5cm
Ted Jacobson\\
{\it Dept. Physics, Univ. of Maryland, College Park, MD 20742-4111}

\end{center}

\begin{abstract}
In this short essay we review the arguments showing that
black hole entropy is, at least in part,
``entanglement entropy", i.e.,
missing information  contained in correlations between quantum field
fluctuations inside and outside the event horizon.
Although the entanglement entropy depends
upon the matter field content of the theory, it turns out
that so does the
Bekenstein-Hawking entropy  $A/4\hbar G_{ren}$,
in precisely the same way, because
the effective gravitational constant
 $G_{ren}$ is renormalized by the very same quantum
fluctuations.
It appears most satisfactory if
the entire gravitational action is ``induced", in the manner
suggested by Sakharov, since then
the black hole entropy is purebred entanglement entropy, rather
than being hybrid with bare gravitational entropy (whatever that
might be.)
\end{abstract}

That the area of a classical black hole event
horizon cannot decrease\cite{areathm} is reminiscent of the
second law of thermodynamics.
Black hole analogs of the zeroth, first, and third laws of
thermodynamics also exist\cite{barcarhaw}, with
the surface gravity $\kappa$ playing the role of temperature.
This analogy gave rise to the idea
that the area is really a measure of entropy\cite{Bek}
It is the behavior of quantum fields in the black hole spacetime
that finally elevates the thermodynamic analogy to an identity.
Hawking's demonstration\cite{Hawk75}
that a black hole in fact radiates quanta at
the temperature $T_H=\hbar \kappa/2\pi$ transforms the surface
gravity
into a true temperature, and fixes the coefficient of the entropy,
$A/4\hbar G$. However, the statistical nature of the black hole
entropy remains to be clarified.

When one thinks of entropy as missing information,
it seems rather natural for an event horizon to have an associated
entropy, since a horizon certainly hides information\cite{Bek}.
But what is the nature of this missing information,
and why does classical general relativity know about it?
In this essay I will recap what is known about this puzzle,
including some very recent developments. I will argue that
all indications point to the following answer: the missing
information is that contained in correlations between quantum field
fluctuations inside and outside the event horizon,
and the reason that classical general relativity knows about this
is that the gravitational dynamics is governed by an action
that is ``induced" by those same quantum fluctuations,
as suggested originally by Sakharov\cite{Sakharov}.

Our starting point is the observation that,
while the Hawking radiation from a black hole appears
thermal from outside the horizon, in fact the global state of the
quantum fields being radiated is pure. It is only when the degrees
of freedom beyond the horizon are ignored that the state appears
as a mixed, thermal one.  I will  call the entropy of this
exterior thermal state  ``entanglement entropy".

The idea  the black hole entropy might arise partly or entirely
from just such entanglement entropy meets with an
immediate difficulty. In continuum quantum
field theory, the entanglement entropy is {\it infinite}, on account of
the singular correlations in the vacuum fluctuations  at pairs of
points on either side of the horizon\cite{SorkinGR10,Sorkin2}.
However, if there is a
physical short distance cutoff on the quantum field degrees of
freedom,
one obtains instead a finite entropy.
Moreover, if the cutoff is chosen
as the Planck length, one gets an entropy of the same order of
magnitude as the Bekenstein-Hawking entropy. Thus, in fact,
one might argue that black hole thermodynamics {\it demands}
such a fundamental cutoff\cite{SorkinGR10,Sorkin2,tHooft}.

This regularization of the divergence in the entropy
appears to lead to another problem, however,
which I will call the ``species problem".
Namely, if the cutoff is fixed at the Planck length $L_{Pl}$, then
the entanglement entropy depends on the number of
independent quantum fields,
whereas the black hole entropy $A/4L_{Pl}^2$
 appears not to.  Of course one could always adjust the
cutoff ``by hand" so as to recover the usual black hole entropy,
however this would be totally artifical.

Different resolutions of the species problem
have been proposed.
It has been suggested that
the whole theory may be  consistent only for a particular spectrum
of physical fields\cite{SorkinGR10,tHooftalk}, or that
the effective cutoff length may, for dynamical reasons,
depend upon the number of
species\cite{SorkinGR10,Sorkin2,FroNov}.
It appears however that the most straightforward resolution
is to keep the fundamental cutoff fixed, but to recognize that
the Planck length itself depends upon the number of species.
While this idea at first sounds mysterious and vague, and seems to
invoke some unknown dynamics of quantum gravity,
 in fact it only requires quantum field theory in a fixed
background spacetime in order to be understood.

The point is that the effective gravitational constant
 $G$ is renormalized by the very same quantum fluctuations
that are giving rise to the entanglement entropy\cite{SussUg}.
It turns out that this renormalization acts in
just the right way to yield the formula $A/4\hbar G_{ren}$
for the black hole entropy, independent of the number of fields
or the form of their interactions.
That is, although the entanglement entropy indeed depends
upon the field content of the theory, so does the
Bekenstein-Hawking entropy, in precisely the same way.

Here I will just sketch the line of reasoning leading to this
conclusion. For simplicity let us assume the black hole is
static (i.e. stationary and nonrotating), and let us
work with the eternal black hole spacetime, i.e. the
maximal analytic extension
(the Kruskal extension in the Schwarzschild case).
The use of the eternal black hole background here
is intended
as an idealization that simplifies the analysis. The significant
conclusions should apply as well to the case of
a black hole that forms from collapse.
The quantum state $\vac$ of the (possibly interacting, generic,)
quantum field $\phi$
is taken to be the unique one that is time-independent and
regular everywhere on the extended spacetime including the horizon.
In the Schwarzschild case this state is the  Hartle-Hawking
``vacuum".

The field degrees of freedom can be partitioned into those
in region $I$, ``outside" the horizon, and those in region $II$,
``inside" the horizon. (For the eternal black hole regions $I$ and $II$
are in fact isometric.)
Tracing over the field degrees of freedom behind the
horizon, we obtain the reduced density matrix
$\rho_I\equiv \tr_{II}|0\rangle\langle0|$, which describes a {\it mixed}
state rather than a pure one, due to the entanglement of
the degrees of freedom in regions $I$ and $II$ in the state
$\vac$.
Rather remarkably, it turns out that this reduced density matrix
can also be expressed as
$\rho_I={\cal N}\exp(-\beta H)$, where
$\beta$ is the inverse Hawking temperature $2\pi/\hbar\k$,
$H$ is the hamiltonian that generates the
time translation symmetry outside the horizon,
and $\cal N$ is a normalization factor.
(Various aspects or cases of this relation have been derived, e.g.,
in \cite{DeWitt,GibbPerry,UnruhWeiss,KabStras,CalWil,BFZ,tj2b}.)
That is, the reduced density matrix is simply proportional
to the density matrix for the canonical thermal ensemble
at the Hawking temperature, defined with repect to the hamiltonian
of the static observers outside the black hole horizon.
In particular, the entanglement entropy is
precisely the same as the thermal entropy
of this ensemble. In terms of the partition function
$Z[\beta]=\tr\exp(-\beta H)$, the entanglement entropy
$-\tr \rho_I\ln\rho_I$ can thus be expressed as the thermal entropy
$(1-{\scriptstyle\beta}\frac{\partial}{\partial\beta})\ln Z$.

To establish the relation between this thermal entropy and the
Bekenstein-Hawking entropy,
one can express the partition function $Z[\beta]$
as a functional integral over
fields on the static Euclidean black hole geometry, with $\hbar\beta$
being the period of the Euclidean time coordinate\cite{SussUg,KabStras}.
The general covariance
of the matter action implies that
the result, $Z=Z[g_{ab}]$, is a generally covariant
functional of the black hole metric $g_{ab}$.
 The effective action $W=-\hbar\ln Z$ has both
local and non-local contributions.
For the local part $W_{loc}$,
a curvature expansion can be written down:
$$W_{loc}=\hbar\int d^4x \sqrt{g}
\Bigl[a_0 +a_1 R + a_2 R^2 + a_2' R^{ab}R_{ab}+ ...\Bigl].$$
The coefficients $a_0$, $a_1$ and $a_2$
generically diverge in four spacetime dimensions. They can be
rendered finite by imposing a short distance
cutoff $L_c$ on the field fluctuations\footnote{If the cutoff breaks
general covariance, then one should no longer expect the effective
action to contain only generally covariant terms. (I thank R.C.~Myers
for pointing this out.) Still, for background metrics
that are slowly varying on the scale
of the cutoff, the generally covariant terms should dominate.},
yielding
$a_0\sim L_c^{-4}$, $a_1\sim L_c^{-2}$, and $a_2\sim \ln L_c$.
We have already seen that such a cutoff is in any case required in
order to avoid an infinite contribution to the entropy.

Now the entanglement entropy receives contributions from
the entire effective action, including the nonlocal parts.
Most of these contributions are state-dependent.
However, the contribution from the most singular part of $W_{loc}$
is universal for states with the same short distance structure.
The cosmological constant term $a_0$ yields no contribution, so the
leading contribution comes from the Einstein-Hilbert term
$\hbar a_1\int d^4x\sqrt{g} R$. Since the metric $g_{ab}$
is time-independent, the volume contribution to the action will
simply be proportional to $\beta$, so will not contribute
to the entropy. In fact, the only contribution  to the entropy
made by $W_{loc}$ comes from the
horizon, where a conical singularity is introduced when $\beta$
is varied away from $2\pi/\hbar\k$. This contribution
is given by $4\pi\hbar a_1 A$,
where $A$ is the area of the horizon\cite{BTZ,SussUg}.
(Note we do not need to assume the black hole is a stationary
point of the effective action to arrive at this conclusion.)
This is of the same form as the Bekenstein-Hawking entropy,
with $\hbar a_1$ in place of $(1/16\pi G)$.\footnote{The
higher curvature terms in $W_{loc}$ yield (state-dependent)
contributions to the entropy that appear to
be naturally asociated with the horizon as well.
A powerful technology for computing these contributions
has recently been developed\cite{noether} and applied\cite{JKM,IyWa}
to a wide
class of higher curvature terms. (See also \cite{Visser} for
similar results obtained by a different method.)}

In other words, the entanglement entropy
of the quantum fields will always contribute to
the renormalized Bekenstein-Hawking entropy
$A/4\hbar G_{ren}$ in proportion to their contribution to
the  renormalized inverse gravitational constant $1/G_{ren}$
in the effective action. This is true independent of their number
and the nature of their interactions.

If the entire gravitational action were induced, then
{\it all} of the black hole entropy
 would be entanglement entropy.
 This would appear much more natural than
if it were a hybrid of entanglement entropy and
``bare gravitational entropy" (whatever that is.)
Whether an induced gravity theory
is a discrete pregeometric dynamics with an intrinsic
cutoff(see, e.g., \cite{thooftradical,bombellietal,Shamir}),
or a superstring
theory (as suggested in this context in \cite{SussUg}), or
something yet entirely different, it receives a
 stamp of approval from black hole thermodynamics and
Occam's razor.

\vskip .5cm
I would like to thank B.L.~Hu and R.C.~Myers for illuminating
discussions and valuable suggestions on the manuscript, and especially
J.Z.~Simon for extensive helpful discussions.
This work was supported by NSF grant PHY91-12240.

\end{document}